**Testing the limits of quantum mechanics**

The physics underlying non-relativistic quantum mechanics can be summed up in two postulates. Postulate 1 is very precise, and says that the wave function of a quantum system evolves according to the Schrodinger equation, which is a linear and deterministic equation. Postulate 2 has an entirely different flavor, and can be roughly stated as follows: when the quantum system interacts with a *classical measuring apparatus*, its wave function collapses, from being in a superposition of the eigenstates of the measured observable, to being in just one of the eigenstates. The outcome of the measurement is random and cannot be predicted; the quantum system collapses to one or the other eigenstates, with a probability that is proportional to the squared modulus of the wave function for that eigenstate. This is the Born probability rule.

Since quantum theory is extremely successful, and not contradicted by any experiment to date, one can simply accept the $2^{nd}$ postulate as such, and let things be. On the other hand, ever since the birth of quantum theory, some physicists have been bothered by this postulate. The following troubling questions arise. How exactly is a *classical* measuring apparatus defined? How large must a quantum system be, before it can be called classical? The Schrodinger equation, which in principle is supposed to apply to all physical systems, whether large or small, does not answer this question. In particular, the equation does not explain why the measuring apparatus, say a pointer, is never seen in a quantum superposition of the two states `pointer to the left' and `pointer to the right'? And if the equation does apply to the (quantum system + apparatus) as a whole, why are the outcomes random? Why does collapse, which apparently violates linear superposition, take place? Where have the probabilities come from, in a deterministic equation (with precise initial conditions), and why do they obey the Born rule? This set of questions generally goes under the name `the quantum measurement problem'. (*Steven Weinberg, The trouble with quantum mechanics, The New York Review of Books, 2017*; *A. J. Leggett, J. Phys.: Condens. Matter 14 (2002) R415–R451*).

For the first sixty years since Schrodinger discovered his equation in 1926, various reinterpretations and mathematical reformulations of quantum mechanics were put forth, to address this problem. They all had one thing in common: they make the same experimental predictions as quantum theory; thus one does not really have concrete evidence to prefer one interpretation over the other. This resulted in endless debates amongst physicists and philosophers who supported one or the other interpretation, a debate which continues to this day. [We note that the experimentally confirmed phenomenon of environmentally induced decoherence does not solve the measurement problem, unless accompanied by a suitable reinterpretation of quantum mechanics. Decoherence destroys interference amongst alternatives, thus converting quantum probabilities into classical ones. But it does not destroy superposition, and hence cannot by itself explain why collapse of the wave function is seen during a measurement].

This situation improved in the 1980s when four physicists, three of them Italian and one American, proposed a falsifiable modification of non-relativistic quantum mechanics, known as Spontaneous Localization (*Phys. Rev. D 34, 470 (1986); A 42, 78 (1990)*). They replaced the second postulate by a precise mathematical postulate, which we label 2'. It states:

The wave function of every particle in nature randomly undergoes spontaneous collapse in position, at a mean rate $\lambda$, and is localized to a size $r_c$. Between every two collapses, the wave function evolves according to the Schrodinger equation.

Postulates 1 and 2' completely define the new theory, which as we shall see shortly, solves the measurement problem, while agreeing with all known tests of quantum mechanics to date. Note that the words `classical measuring apparatus' no longer appear in the definition of the theory. Here, $\lambda$ and $r_c$ are two new constants of nature, for which the authors proposed the values $\lambda \approx 10^{-16} \, s^{-1}$ and $r_c \approx 10^{-5}$ cm. These are the most conservative choices which allow a solution of the measurement problem, but eventually the values of these two new constants must be determined by experiment. This value of $\lambda$ is assumed to be the collapse rate for a nucleon, so that the `particle' in the above postulate 2' is actually assumed to be a nucleon. What this means is that if one creates a superposed state `proton here + proton there', this state will on the average collapse only once in $10^{16}$ s, which is an astronomical time scale, and this explains why we never see a proton wave function collapse on its own. Quantum linear superposition is obeyed here to a very high accuracy, and we see interference fringes in a double slit experiment with protons.

Consider next a bound system of two nucleons, say a deuteron, and a superposition of two position states of the deuteron: `deuteron here' + `deuteron there'. More explicitly, it is the state [ (proton here) * (neutron here) + (proton there) * (neutron there) ]. Such a state is called an entangled state. It is evident that spontaneous collapse of either the proton or the neutron will localize the deuteron, and hence the collapse rate of the deuteron is 2$\lambda$. Now if we have a macroscopic object consisting of N nucleons, the collapse rate will be N$\lambda$. Since N is enormous for a large object, spontaneous localization will take place extremely rapidly: in less than a millionth of a second, for N$\approx 10^{23}$. That is why we never see, a chair say, in a superposed state such as `chair here' + `chair there'. Position superpositions of macroscopic objects collapse extremely rapidly. The quantum superposition principle thus becomes an approximate principle, holding for very large times for microscopic systems, but for extremely small times, for macroscopic ones. Through this elegant amplification mechanism, spontaneous localization provides a unified description of microscopic and macroscopic dynamics (*Science, 325 275 (2009)*).

The measurement problem is solved as follows. Suppose the incoming quantum system is an electron in a superposition of two spin states: (spin up + spin down). The interaction with the measuring apparatus creates an entangled state of the type [(spin up)*(pointer to the left) + (spin down)*(pointer to the right)]. Since the pointer is a macroscopic object, the superposition of its left and right states spontaneously collapses extremely rapidly, and the entangled state is quickly reduced to either (spin up)*(pointer to the left) or (spin down)*(pointer to the right). This is interpreted as the collapse of the wave function. It can be proved that the outcome of the collapse is randomly selected, and that it obeys the Born probability rule.

Macroscopic quantum systems such as superconductors, superfluids and Bose-Einstein condensates are long-lived, because they are described by product states, not by entangled states. To take the example of two nucleons, these are states of the form (proton here + proton there)*(neutron here + neutron there). Here, the two nucleons are not bound, and spontaneous collapse of one does not collapse the other. The collapse rate of such a state is $\lambda$, not $2\lambda$. The amplification mechanism does not work for such states. On the other hand, while entangled states of BECs have indeed been made in the laboratory, they involve only a few tens of nucleons, and hence are long lasting.

It is natural to take a continuum limit of spontaneous localization, and combine the two postulates 1 and 2' into a unified mathematical description of dynamics. This is done by modifying the Schrodinger equation into a stochastic nonlinear differential equation. The form of this equation is strongly constrained by the requirement that faster than light signaling is not allowed, and that the collapse of the wave function should obey the Born probability rule. In this theory, known as Continuous Spontaneous Localization (CSL), the linear dispersive aspect of the Schrodinger evolution, and the nonlinear localizing aspect of spontaneous collapse, compete with each other. The collapse rate is proportional to mass: for a particle of mass $m$ it is $m\lambda/m_N$ where $m_N$ is the nucleon mass. Localization is assumed to be caused by a universal noise field, whose origin remains to be ascertained. It could be a yet to be discovered field of gravitational or cosmological origin perhaps, or a relic of a deeper theory underlying quantum mechanics (*Reviews of Modern Physics, 85, 471 (2013)*). One consequence of the interaction of a quantum system with the noise field is that the system gains energy from the noise field, at an extremely tiny rate, proportional to the collapse rate $\lambda$, which induces a random walk in the motion of a particle, or a bulk heating in a solid (*Phys. Rev. Lett.73, 6 (1994), Phys. Rev. A 97, 052119 (2018)*).

The introduction of the two new constants $\lambda$ and $r_c$ makes the experimental predictions of CSL different from those of quantum mechanics. For microscopic systems, the difference is impossibly small to detect, but the deviation from quantum mechanics can become detectable for mesoscopic and macroscopic systems. The most direct test is to verify the presence of quantum linear superposition in an interference experiment. Of course, interference has been observed for micro-systems such as electrons, neutrons, atoms, and ions. Then there was the celebrated Vienna experiment which observed interference in fullerene molecules $C_{60}$ (*Nature 401, 680 (1999)*). CSL predicts that if a beam of sufficiently heavy particles (made of about a billion nucleons) is sent through a double slit or a diffraction grating, the resulting superposed state will collapse before reaching the screen, and an interference pattern will not be seen. The largest composite objects for which interference has been tested and confirmed are made up of about ten thousand nucleons, and this observation sets an upper bound of about $10^{-5}$ $s^{-1}$ on $\lambda$ (any larger and collapse would happen too fast and destroy the interference pattern). Doing such experiments for even larger objects poses extraordinary technological challenges as regards beam preparation, and maintaining spatial and temporal coherence (*Nature Physics 10, 271 (2014)*, *Rev. Mod. Phys. 84, 157 (2012)*). Useful constraints have come also from cold atom interferometry. Optomechanical experiments have been attempted as well, to create superposition states of macroscopic objects. It has been proposed that to improve the interferometry experiments, it will be necessary to go a micro-gravity environment in outer space (www.qtspace.eu), (maqro-mission.org).

Important upper bounds on the collapse rate come from already known physics. The tiny energy gained from the collapse inducing noise field should not cause so much random motion or heating that it contradicts known physical phenomena. This provides interesting limits from astrophysics and cosmology, from spontaneous emission, and from decay of supercurrents, amongst others (*J. Phys. A 40, 2935 (2007), arXiv:1807.11450*). Some of the strongest upper bounds have come from the observed stability of the LISA pathfinder (*Phys. Rev. D **94**, 124036 (2016)*) and from spontaneous X-ray emission from free electrons in solid Germanium (*J. Adv. Phys. 4 (2015) 263*). There is a theoretical lower bound on $\lambda$ and there is a theoretical bound on the range of $r_c$, if CSL is to solve the measurement problem. Based on these observations one can make an exclusion diagram in the $\lambda - r_c$ plane, and the currently allowed values are boxed in from all four sides, leaving a closed allowed region in the middle, which spans about eight orders of magnitude in $\lambda$ and a few orders around $10^{-5}$ cm for $r_c$ (*arXiv:1805.10100 and references therein*). (Quantum mechanics is the limit $\lambda \to 0$ and $r_c \to \infty$). The present experimental upper bound on $\lambda$ stands around $10^{-8} s^{-1}$.

Very significant developments started taking place around 2015, and the experimental and theoretical scene has made impressive progress since then. Following older work, we showed that at currently achievable low temperatures and pressures, it is possible to detect the CSL induced random motion of a micron sized particle (*Sci. Rep. 5, 7664 (2015)*). This led to a series of papers proposing the so-called non-interferometric tests of CSL, which all aim to detect effects of the energy gain from the CSL noise field. This culminated in an experiment which looks for the CSL noise effect in an ultra-cold micro-cantilever and reported that *``high accuracy measurements of the cantilever thermal fluctuations reveal a nonthermal force noise of unknown origin''* (*Phys. Rev. Lett. **119**, 110401 (2017)*). This noise is consistent with being a CSL effect with $\lambda \cong 10^{-8}$ and constitutes the first reported detection, which the community has received with caution. A repeat of this experiment at a different laboratory is currently in planning stages. Furthermore, the European Commission, as a part of its `Horizon 2020 Research and Innovation programme' has funded a promising dedicated experiment, TEQ, which will look for this random motion using a levitated nanosphere (tequantum.eu). TEQ involves a collaboration between eight European research organisations over the next three years, and the experiment is expected to significantly improve the current bounds on the collapse rate, or actually detect CSL. The nanosphere will be levitated by employing external electric fields, and its random motion will be looked for via deflection of incident laser beams.

Another promising experimental proposal is the bulk heating of a solid through phononic excitation modes which couple to the CSL noise field (*Phys. Rev. A A 97, 052119 (2018)*). If the experiment is carried out at ultra-low milli-Kelvin temperatures, where the specific heat is extremely low, it will in principle be possible to detect a very tiny increase in the solid's temperature, which is being caused by CSL noise. It turns out that the extent of heating is low enough that it competes with heating of the solid via incident cosmic rays and ambient radioactivity. If the experiment is carried out in an underground physics laboratory, it will perhaps be possible to probe the lowest permitted $\lambda$ value of $10^{-16} s^{-1}$ (*arXiv:1807.03067*). One can expect some exciting and significant progress in the next few years, as experiments gear up to confirm or rule out the theory of spontaneous localization.

There are several outstanding theoretical challenges as well, for CSL. Foremost is to understand the origin of the CSL noise field. Also, constructing a relativistic version of CSL has proven to be a great challenge. There is at present a debate in the community as to whether it is at all possible to make such a version, or whether spontaneous localization will force us to revise our ideas about special relativity and space-time structure. There is likely a connection between the problem of time in quantum theory, and the measurement probem. (arXiv:1210.8110). It has also recently been suggested that space, and perhaps time as well, themselves emerge as a consequence of collapse of the wave-function (*arXiv:1806.01297*). In the coming few years, we can hope to learn whether quantum theory is exact, or an approximaton to a deeper theory. Either way, we will have made significant progress in our understanding of quantum physics.

Tejinder Pal Singh
Tata Institute of Fundamental Research
Homi Bhabha Road, Mumbai 400005, India
e-mail: tpsingh@tifr.res.in